\documentclass[10pt,A4paper,conference]{IEEEtran}
\IEEEoverridecommandlockouts
\usepackage{cite}
\usepackage{amsmath,amssymb,amsfonts}
\usepackage{graphicx}
\usepackage{textcomp}
\usepackage{xcolor}
\usepackage{amsmath}
\usepackage[misc]{ifsym}

\usepackage{tabularx}
\usepackage{color, colortbl}
\usepackage[misc]{ifsym}
\usepackage{xcolor}
\definecolor{lime}{rgb}{0.88,2,10}

\newcommand*{\Resize}[2]{\resizebox{#1}{!}{$#2$}}%

\renewcommand{\baselinestretch}{1.05}
\usepackage{subcaption}
\usepackage{wrapfig}
\usepackage[linesnumbered,ruled,vlined]{algorithm2e}
\usepackage{xpatch}
\usepackage{url}
\usepackage{multirow}
\usepackage{multicol}
\usepackage{wrapfig}

\usepackage[export]{adjustbox}
\usepackage{nomencl}
\usepackage{siunitx}
\usepackage{blindtext}
\usepackage[T1]{fontenc}
\usepackage[spaces,hyphens]{xurl}
\usepackage{float}
\usepackage{comment}

\usepackage{diagbox}

\SetAlFnt{\small}

\SetKwComment{Comment}{$\triangleright$\ }{}

\usepackage[utf8]{inputenc}
\usepackage{enumitem}

\SetKwInput{kwInit}{Initialize}

\usepackage{booktabs}

\def\BibTeX{{\rm B\kern-.05em{\sc i\kern-.025em b}\kern-.08em
    T\kern-.1667em\lower.7ex\hbox{E}\kern-.125emX}}

\usepackage[numbers,sort&compress]{natbib}
\usepackage[font={small}]{caption} 

\newcommand{\fref}[1]{Fig.~\ref{#1}}
\newcommand{\tref}[1]{Table~\ref{#1}}
\newcommand{\sref}[1]{Section~\ref{#1}}
\usepackage[misc]{ifsym}

\usepackage{tikz}

\newcounter{stepnum}

\usepackage[outline]{contour}
\contourlength{0.2pt}
\contournumber{15}

\newcommand\HUGE{\fontsize{20.74}{25}\selectfont}

\begin{document}

\title{\HUGE{Efficient Brain Imaging Analysis for Alzheimer's and Dementia Detection Using Convolution-Derivative Operations}}

\author{\IEEEauthorblockN{
Yasmine Mustafa\IEEEauthorrefmark{1}, Mohamed Elmahallawy\IEEEauthorrefmark{2}, Tie Luo\IEEEauthorrefmark{3} \textsuperscript{\Letter}\thanks{\textsuperscript{\Letter} Corresponding author.}}  
\IEEEauthorblockA{%
 \IEEEauthorrefmark{1}Computer Science Department, Missouri University of Science and Technology, Rolla, MO 65401, USA}
\IEEEauthorblockA{%
 \IEEEauthorrefmark{2} Department of Computer Science and Cybersecurity, Washington State University, Richland, WA 99354, USA}
\IEEEauthorblockA{%
 \IEEEauthorrefmark{3}  Department of Electrical and Computer Engineering, University of Kentucky, Lexington, KY 40506, USA }
Emails:  yam64@mst.edu, mohamed.elmahallawy@wsu.edu, t.luo@uky.edu}
\maketitle

\begin{abstract}
Alzheimer's disease (AD) is characterized by progressive neurodegeneration and results in detrimental structural changes in human brains. Detecting these changes is crucial for early diagnosis and timely intervention of disease progression. Jacobian maps, derived from spatial normalization in voxel-based morphometry (VBM), have been instrumental in interpreting volume alterations associated with AD. However, the computational cost of generating Jacobian maps limits its clinical adoption. In this study, we explore alternative methods and propose {\em Sobel kernel angle difference} (SKAD) as a computationally efficient alternative. SKAD is a derivative operation that offers an optimized approach to quantifying volumetric alterations through localized analysis of the gradients. By efficiently extracting gradient amplitude changes at critical spatial regions, this derivative operation captures regional volume variations Evaluation of SKAD over various medical datasets demonstrates that it is 6.3$\times$ faster than Jacobian maps while still maintaining comparable accuracy. This makes it an efficient and competitive approach in neuroimaging research and clinical practice.
\end{abstract}


\section{Introduction}
\begin{figure*}[t]
    \centering
     {\subfloat[Original grid.]{\includegraphics[width=0.293\textwidth]{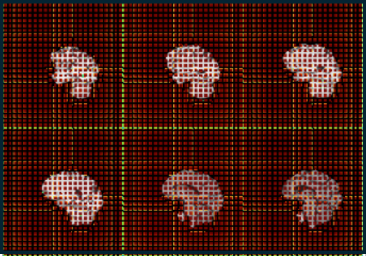}}}
    {\subfloat[Warped grid of a cognitively normal subject.]{\includegraphics[width=0.33\textwidth]{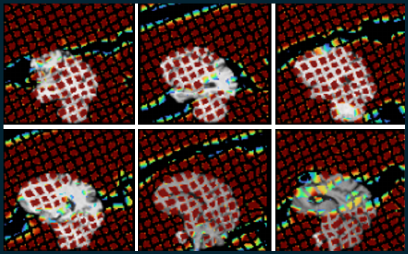}}}
    {\subfloat[Warped grid of a subject with dementia.]{\includegraphics[width=0.34\textwidth]{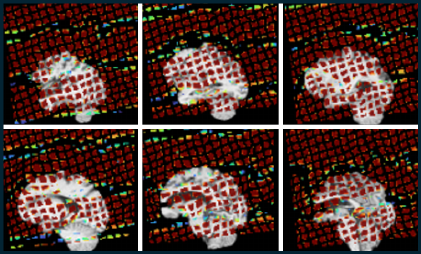}}}
\caption{Visualization of a deformation field from 6 view angles through deforming overlaid grids.}  \label{fig:grid}
\end{figure*}
Dementia is a major cause of disability and dependence among aging populations around the world. About 35.6 million people were diagnosed with dementia in 2010 and, more alarmingly, this number is estimated to rise to 65.7 million by 2030 and 115.4 million by 2050 \cite{prince2013global}. Alzheimer's disease is the most common form of dementia, accounting for approximately 60-80\% of all dementia cases. As the discussion throughout this paper applies to both, we will use them interchangeably. Dementia profoundly impacts cognitive and physical abilities, hindering everyday tasks, placing a significant emotional and financial burden on families, and escalating healthcare costs. Unfortunately, there is currently no effective cure for late-stage dementia. As a result, early detection and intervention are critical. This task, however, is challenging because pre-symptomatic changes in the brain are extremely subtle, yet they represent a crucial window of opportunity for timely intervention. Therefore, developing effective early detection methods remains an urgent and unmet need.

Structural magnetic resonance imaging (MRI) plays a critical role in detecting brain tissue loss, such as atrophy, which is caused by pathological changes associated with neurodegenerative disorders. Distinct patterns of tissue loss are often linked to specific stages of dementia, typically categorized into four stages: mild cognitive impairment (MCI), mild dementia, moderate dementia, and severe dementia. Among these, the MCI and mild stages are particularly significant for early intervention but remain challenging to detect.

Traditionally, analyzing atrophy on MRI has relied on visual assessments by experienced radiologists and manual measurements. Recently, computer-aided diagnostic (CAD) tools have gained significant attention by leveraging advanced imaging techniques and machine learning. A prominent CAD method is voxel-based morphometry (VBM), which performs statistical computations across all voxels in a 3D image to detect anatomical differences in the brain among subjects or groups. These differences can highlight areas affected by atrophy or tissue expansion caused by brain diseases \cite{whitwell2009voxel}.

Preprocessing is a critical step in brain imaging analysis, including VBM. One such preprocessing technique is the Jacobian map, a powerful tool for quantifying volumetric changes in the brain. It is particularly useful in comparing Alzheimer's patients to healthy controls. Several studies have successfully employed Jacobian maps as inputs for machine learning models to identify regions of brain atrophy and achieved promising results \cite{spasov2019parameter,abbas2023transformed,mustafa2023diagnosing,mustafa2024unmasking}.

However, the computation of Jacobian maps is highly resource-intensive. This challenge becomes particularly pronounced in large-scale image analyses involving thousands of scans, significantly limiting their clinical adoption. Furthermore, with the growing prevalence of multi-modal healthcare models, many of which incorporate vision, efficient image preprocessing methods are becoming increasingly vital to ensure scalability and practical applicability. Moreover, large pretrained models are increasingly being applied to healthcare research, but they require vast quantities of images; simultaneously, new approaches to assembling extensive medical datasets, such as leveraging textbooks, are emerging. Consequently, large pretrained models will also likely benefit from efficient medical image preprocessing techniques, which significantly reduce these models' training time and resource demands, thereby making them more practical and scalable for healthcare applications.

In this paper, we propose an efficient image preprocessing technique for quantifying volumetric changes in the brain, for more scalable neuroimaging research and practices. Specifically, our contributions are as follows:
\begin{itemize}
    \item We introduce the \textit{Sobel Kernel Angle Difference} (SKAD) method to analyze volumetric alterations associated with Alzheimer’s disease and other types of dementia, and potentially other neurodegenerative disorders such as Parkinson’s disease. This approach achieves more than 6 times computational efficiency compared to Jacobian maps, the current state-of-the-art technique, while delivering competitive accuracy.
    
    \item Our extensive evaluations and ablation studies demonstrate that SKAD effectively fulfills our objectives, reducing computation time by 84\% (from 584 ms to 93 ms) and floating-point operations by a remarkable 99\% (from 146 billion to 1.66 billion).
\end{itemize}

\section{Related Work}
Incorporating Jacobian determinants in image preprocessing dates back to the early 2000s \cite{ashburner2000voxel}. Wright et al. \cite{wright1995voxel} introduced a novel technique to characterize regional cerebral grey and white matter differences in MRI, where images are segmented and then spatially normalized to a symmetrical template brain in stereotactic Talairach space. Spatial normalization, commonly known as \textit{nonlinear registration}, is a process that involves transforming all the data of a subject to a common stereotactic space, allowing for a more accurate and consistent data analysis. Certain brain regions experience volume expansion during the nonlinear registration while others undergo contraction. This phenomenon complicates the interpretation of VBM results, which aim to identify regional differences in tissue concentration. To address this issue, Ashburner et al. \cite{ashburner2000voxel} utilized the Jacobian determinants to represent the relative voxel volumes. By multiplying segmented images by these determinants, the actual amounts of grey matter within each structure are preserved, allowing for a more accurate comparison of grey matter concentrations across regions. This process is referred to as "Jacobian modulation" and was optimized by Good et al. \cite{good2001voxel}. Currently, it is employed by the FMRIB Software Library (FSL), a widely used software library for brain image analysis. \newline
One of the improvements of VBM is tensor-based morphometry (TBM), which has similar steps to VBM but does not require segmentation, hence avoiding the issue of accurate tissue classification \cite{kim2008structural}. TBM has been used to detect group differences between healthy controls and various patient populations. This attracted researchers to analyze each medical image as a whole instead of segmenting it. In VBM,  Jacobian determinants are used in the modulation step as correction for the registration step. More recently, the Jacobian determinant values are visualized across the image or regions of interest to create the \textit{Jacobian map}, which is then used as the main input to the training models \cite{mustafa2024unmasking,abbas2023transformed,mustafa2023diagnosing,spasov2019parameter}. This map provides a spatial representation of the local volume changes occurring throughout the image, allowing for qualitative and quantitative analysis of deformation patterns. Recently, deep learning techniques, like convolutional neural networks (CNNs), have been proposed to diagnose Alzheimer's disease (AD) by learning atrophy patterns in sMRIs. These CNN-based approaches require the identification of discriminative landmark locations by detecting regions of interest in sMRIs. Therefore, the performance of the entire framework is highly dependent on the accuracy of the landmark detection step, and as a result, highlighting areas of deformation is crucial before training the CNN. Hence, the choice of the method that highlights the landmarks is crucial for effectively capturing pathological changes indicative of AD. Abbas et al. \cite{abbas2023transformed} proposed a 3D Jacobian domain CNN (JD-CNN) to diagnose AD subjects.  Utilizing the complete 3D Jacobian Map obviates the necessity of partitioning images into patches, as done in patch-based or region-of-interest (ROI)-based models, thereby enabling the model to learn spatial correlations across different brain regions more effectively. Spasov et al. \cite{spasov2019parameter} utilized Jacobian maps to develop a multi-modal framework aimed at quantifying local volumetric transitions linked to AD.
Jacobian maps have also proved effective in other applications such as motion-tracking cardiac tagging MRI \cite{ye2021deeptag}. Mustafa et al. \cite{mustafa2023diagnosing} used Jacobian maps to develop a multimodal model consisting of random forests (RF) and CNN that trains on the Jacobian maps of MRI and CT scans. In another study conducted by Mustafa et al. \cite{mustafa2024unmasking}, Jacobian maps were used for explainability to improve the trustworthiness of artificial intelligence (AI). Since these maps contain information about the volumetric changes that occur in the brain when compared to a healthy brain template, researchers utilized the Jacobian maps to develop a model-debugging system that assists the CNN model in identifying important areas for detecting dementia during training. 
While the Jacobian maps hold significant importance in medical imaging, particularly in tasks like image registration and deformation analysis, their calculation can be resource-demanding, especially for high-resolution or large-scale image datasets. This computational burden may limit the utility of Jacobian-based analyses in practical clinical or research settings, where timely analysis and interpretation of imaging data are often pivotal. To this end, we introduce Sobel kernels, a specific convolution kernel used for edge detection, as an alternative efficient tool that can also provide insights into brain volumetric changes. Sobel kernels were first introduced as two-dimensional (2D) kernels that work with 2D images \cite{kanopoulos1988design, hussain2014underwater} and were extended to 3D in multiple works \cite{aqrawi2011detecting,prabhakar2024vit}.

\begin{figure*}[t]
  \centering
 \includegraphics[width=\linewidth]{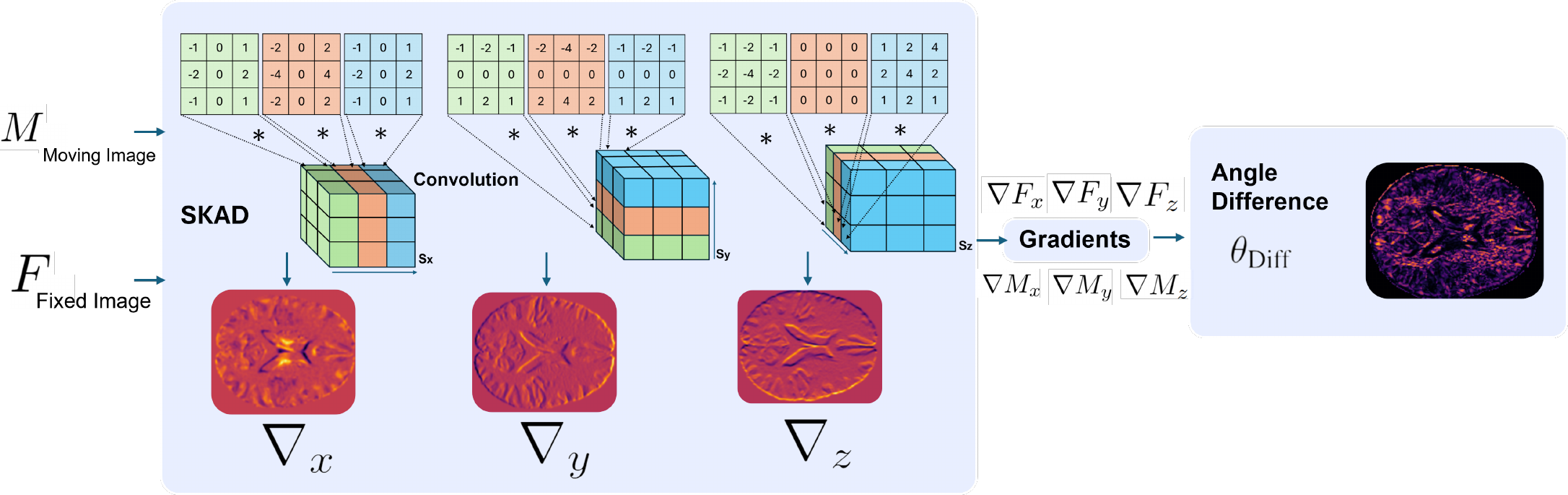} 
  \caption{Schematic illustration of the 3D 3$\times$3 Sobel kernels used in this work. These kernels are applied to both the moving and fixed images to perform convolutions in the x, y, and z directions. The resulting gradients are then used to compute the 3D angle difference.}
  \label{fig:sobel}
\end{figure*}

\section{Methodology}

\subsection{Registration}\label{sec:reg}
Registration is a crucial step for calculating the Jacobian map, enabling spatial alignment and standardization with a healthy template. Spatial alignment ensures that the MRI scans of different subjects are matched accurately to the same anatomical locations, which is essential for comparative analysis and identifying structural changes or abnormalities across individuals. The second purpose, standardization with a healthy template, involves contrasting each individual MRI scan against a reference, so that deviations from typical brain morphology can be assessed objectively and thus help identify pathological changes associated with AD. We used the Montreal Neurological Institute (MNI) 152 standard-space T1-weighted average structural template, which is derived from iteratively registering 152 healthy structural images to the MNI coordinate system and then averaging them all together. The MNI coordinate system is based on a 3D Cartesian coordinate system where the origin is defined at the intersection of the anterior and posterior commissures of the brain \cite{evans2012brain}. The registration process is also computationally and memory intensive, and much effort has been made to improve its efficiency while balancing performance \cite{meng2024correlation}. However, our work specifically targets enhancing the efficiency of calculating volumetric changes and exploring alternatives to Jacobian Maps.

Employing ANTs \cite{avants2009advanced}, we perform nonlinear registration by first calculating a deformation field $V$ of each MRI image and then transforming the image to the MNI 152 template. Unlike linear transformation which involves only translation, rotation, and scaling, nonlinear registration captures more nuanced deformations such as local shape variations and distortions. The deformation field $V$ essentially provides a map of how each voxel in the source image should be displaced or deformed to accurately match its counterpart in the target image (i.e., the MNI template). \fref{fig:grid} provides a visualization where we overlay the deformation map onto fine-grained grids superimposed on MRI images, in order to highlight the extent of warping undergone by the grids. The rationale is that grids corresponding to demented subjects will exhibit more pronounced warping compared to those of normal subjects. This observation also underscores the subtle yet discernible differences between normal brain morphology and deviations present in pathological conditions.
The source image to be warped (i.e., transformed) is also commonly referred to in medical imaging as the \textit{moving image}, $M(x,y,z)$, and the target (i.e., the template) as the \textit{fixed image}, $F(x,y,z)$. Here $x$, $y$, and $z$ represent the three spatial dimensions of the image, corresponding to the coordinates along the width, height, and depth axes, respectively, of a 3D image volume. 

After registration, to highlight the volumetric changes using the warped grids, the current method is to compute the Jacobian map described next. However, we introduce an alternative, SKAD, in \sref{sec:skad}.

\subsection{Jacobian Maps}
A Jacobian map is a matrix $J$ computed from the deformation field $V$ obtained from the above nonlinear registration step. Computing this matrix begins with calculating the first derivative of each voxel $v(x,y,z)$ to encode local deformations, i.e.,
\begin{eqnarray}
J(v)=
\resizebox{!}{1cm}{$
\begin{bmatrix}
\frac{\partial v_x}{\partial x} &  \frac{\partial v_x}{\partial y} &  \frac{\partial v_x}{\partial z} \\
 \frac{\partial v_y}{\partial x} &  \frac{\partial v_y}{\partial y} &  \frac{\partial v_y}{\partial z} \\
 \frac{\partial v_z}{\partial x} &  \frac{\partial v_z}{\partial y} &  \frac{\partial v_z}{\partial z}
\end{bmatrix}
$}
\end{eqnarray}

Then, we calculate the determinant $Det(J)$ for every voxel $v$ which forms the Jacobian map of the moving image $M$:
\begin{eqnarray}
\Resize{7.5cm}{J(M)= \begin{bmatrix}
& \vdots &\\
\dots & \det(J(v(x,y,z))) & \dots \\
& \vdots &\\
\end{bmatrix}_{\begin{matrix}  x=1...W\ (\text{width}) \\   y=1...H\ (\text{height}) \\   z=1...D\ (\text{depth}) \end{matrix}}}, 
\\ \notag
\Resize{7.4cm}{\text{At each voxel: }
\begin{cases}
\text{Volume expansion} & \text{if } \det(J) > 1 \\
\text{No change} & \text{if } \det(J) = 1 \\
\text{Volume compression} & \text{if } \det(J) < 1 
\end{cases}}
\label{volume_changes}
\end{eqnarray}

Hence, the Jacobian map helps identify the volumetric change (as a ratio) of the brain at the voxel level before and after deformation.

\begin{figure*}[t]
    \centering
     {\subfloat[Normal.]{\includegraphics[width=0.46\linewidth]{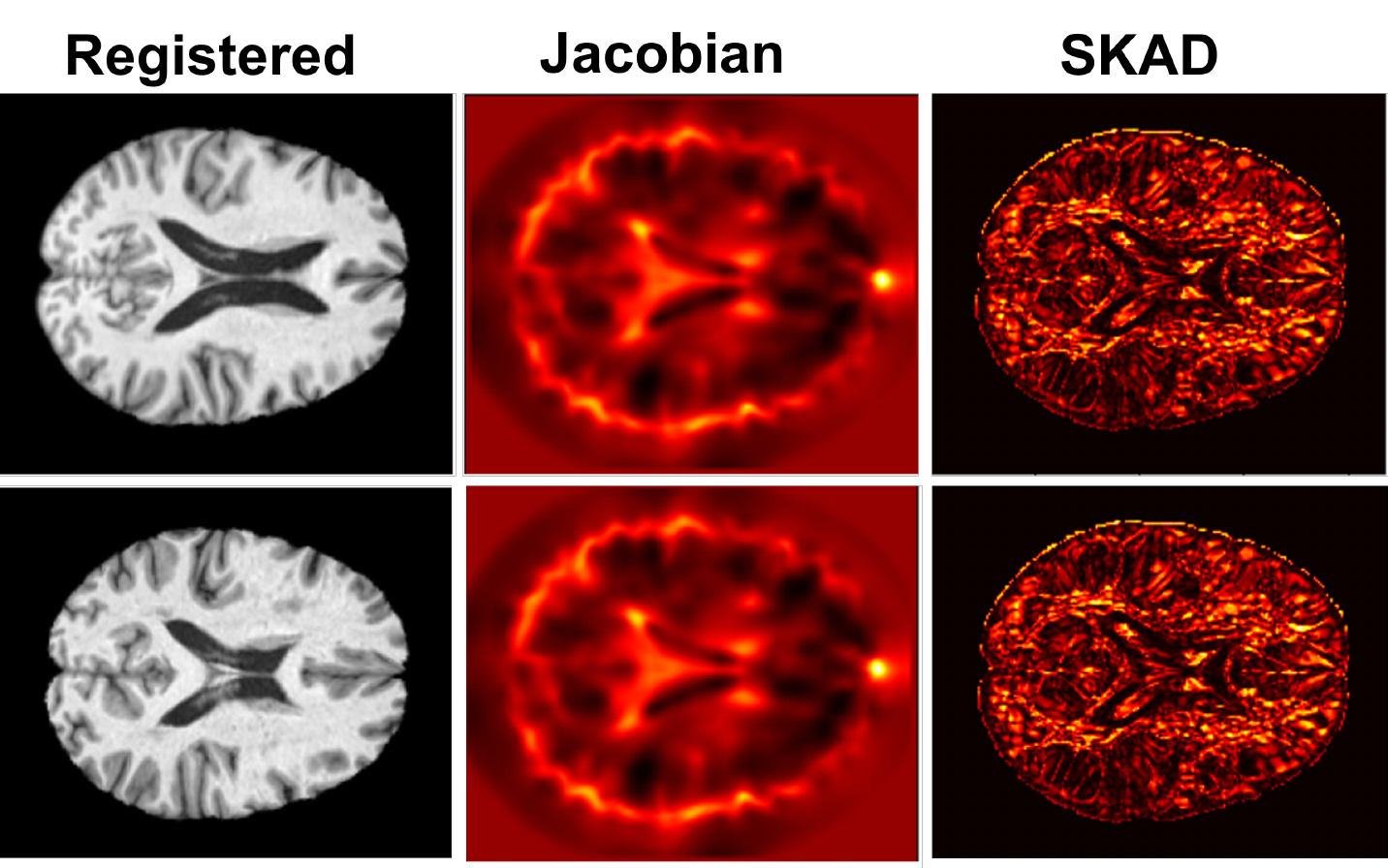}}}
    {\subfloat[MCI.]{\includegraphics[width=0.46\linewidth]{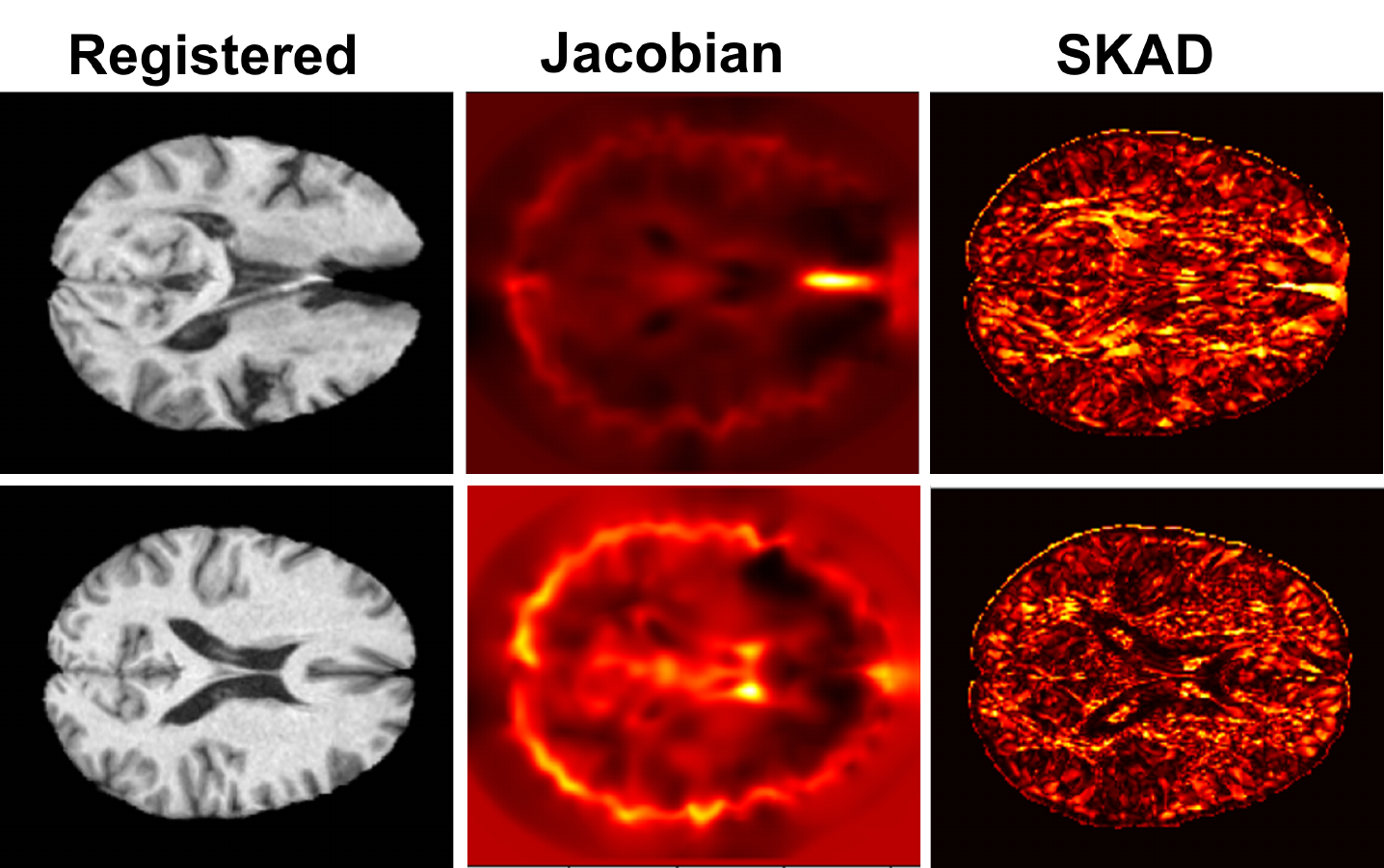}}}

     {\subfloat[Mild Dementia.]{\includegraphics[width=0.46\linewidth]{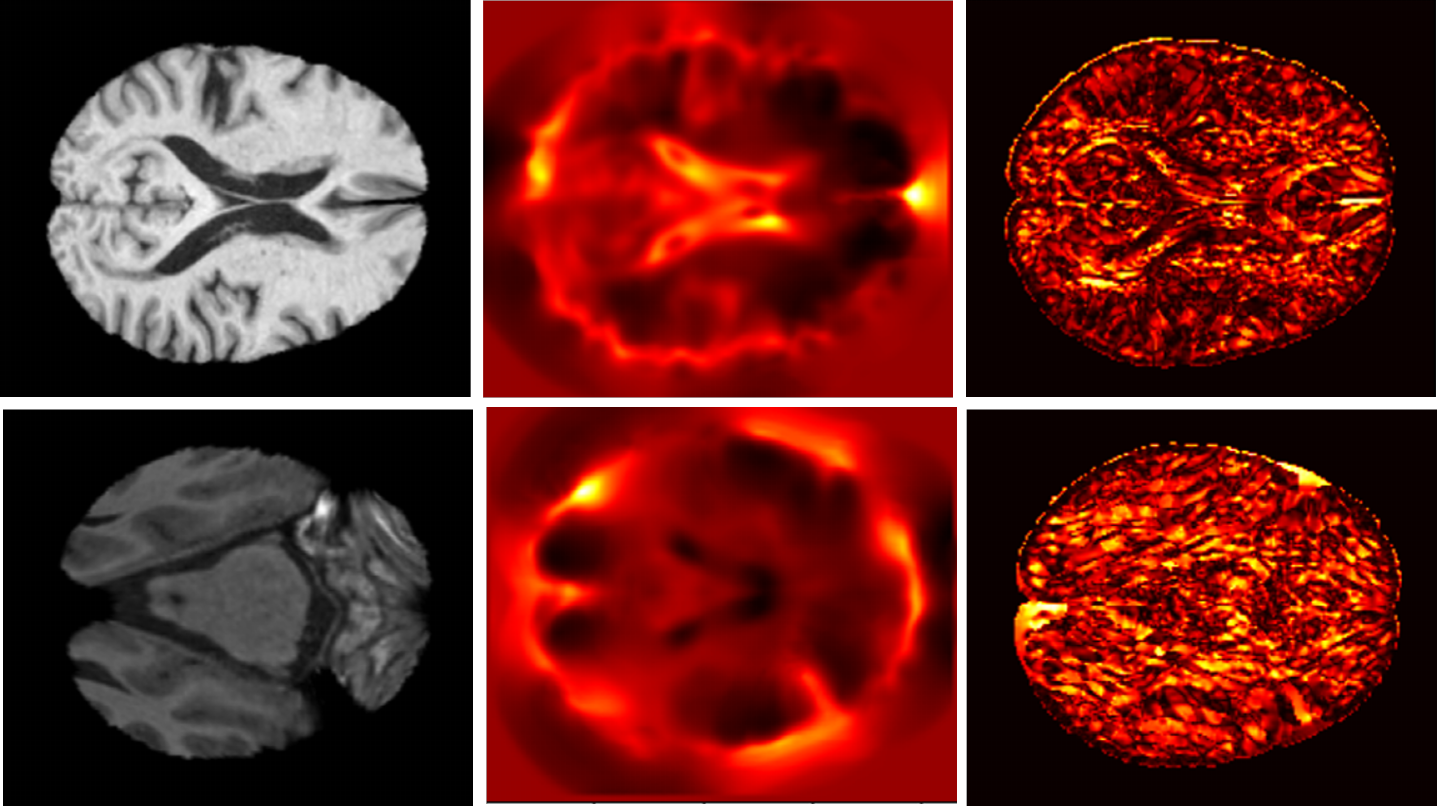}}}
     {\subfloat[Moderate Dementia.]{\includegraphics[width=0.46\linewidth]{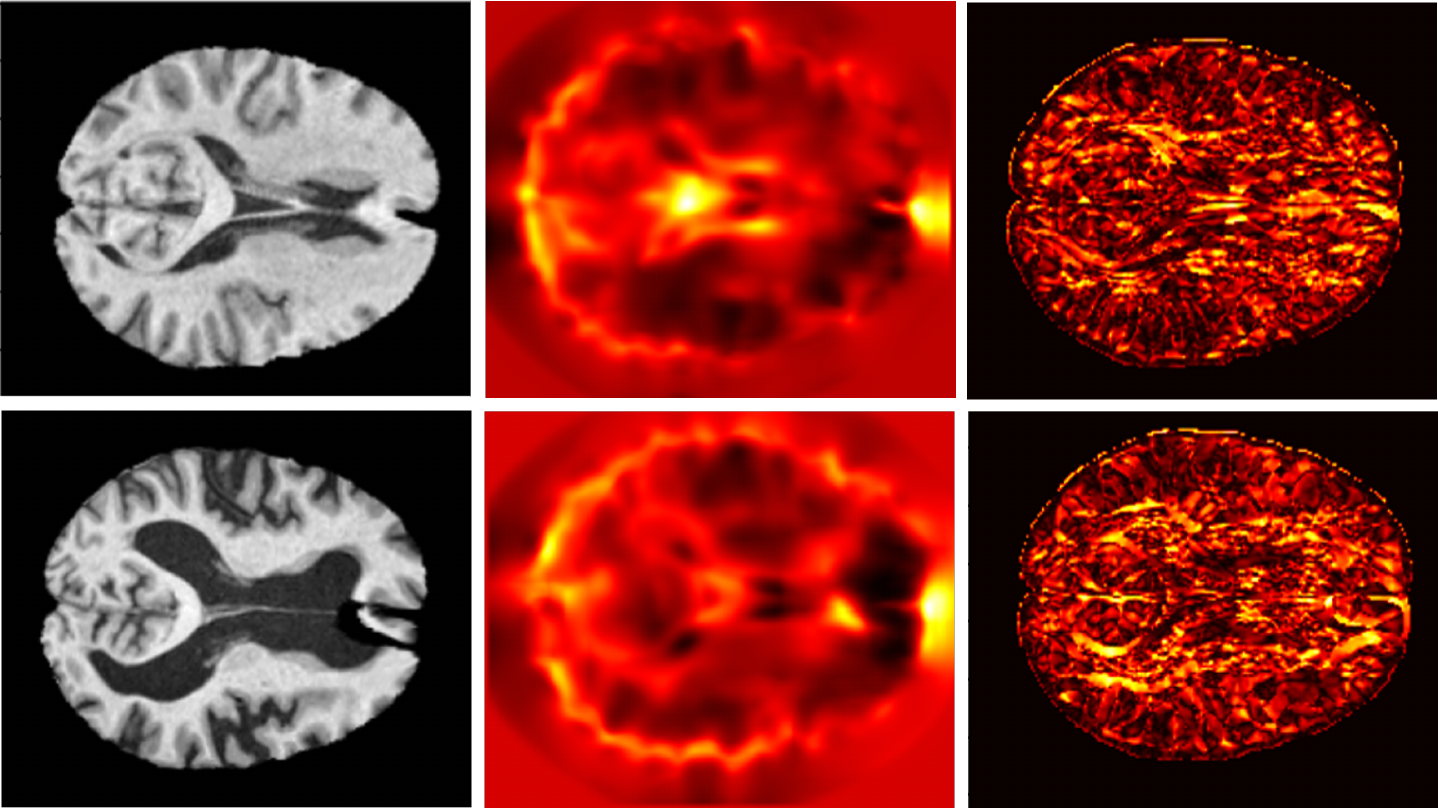}}}
\caption{Visualization of MRI-derived heat maps depicting the scans after registration, Jacobian Maps, and SKAD. Four cognitive stages are shown.} \label{pics}
\end{figure*}

\subsection{Sobel Kernel Angle Difference (SKAD)}\label{sec:skad}
We propose Sobel Kernel Angle Difference (SKAD) as a novel method for characterizing brain volumetric changes. SKAD offers an alternative approach to traditional derivative operations by using convolution, which can be more robust in image processing tasks. Typically, a derivative operation is used to measure the rate of change in a function. In digital image processing, this is often approximated using finite differences. The general form of a first-order derivative in a discrete setting is given by \cite{adetokunbo20163d, gonzalez2008digital}:
\begin{equation}
\frac{\partial f}{\partial z} \approx \lim_{\Delta z \rightarrow 0} \frac{f(x, y, z + \Delta z) - f(x, y, z - \Delta z)}{2 \Delta z},
\end{equation}
where $\frac{\partial f}{\partial z}$ represents the first-order partial derivative with respect to $z$, and $f(x, y, z)$ denotes a 3D input function. This traditional method involves direct computation of the derivatives, which can be sensitive to noise and variations in the data.

On the other hand, Sobel kernels provide a different way of computing gradients without explicit derivative operations, making it an efficient approach for image processing and offering a more precise representation of 3D gradients. By employing convolution with Sobel kernels, SKAD provides a more stable estimation of gradients, preserves spatial relationships, and enhances the detection of structural changes in brain images.

A 3$\times$3$\times$3 Sobel kernel is defined as follows for each direction:

\begin{eqnarray}
    S_x =
    \resizebox{7.5cm}{!}{$
    \begin{bmatrix}
    \begin{bmatrix}
    -1 & 0 & 1 \\
    -2 & 0 & 2 \\
    -1 & 0 & 1 \\
    \end{bmatrix}_{[-1,:,:]}, 
    \begin{bmatrix}
    -2 & 0 & 2 \\
    -4 & 0 & 4 \\
    -2 & 0 & 2 \\
    \end{bmatrix}_{[0,:,:]}, 
    \begin{bmatrix}
    -1 & 0 & 1 \\
    -2 & 0 & 2 \\
    -1 & 0 & 1 \\
    \end{bmatrix}_{[1,:,:]}
    \end{bmatrix}\nonumber
    $}
\end{eqnarray}

\begin{eqnarray}
S_y = 
    \resizebox{7.5cm}{!}{$
\begin{bmatrix}
\begin{bmatrix}
-1 & -2 & -1 \\
 0 &  0 &  0 \\
 1 &  2 &  1 \\
\end{bmatrix}_{[:,-1,:]}, 
\begin{bmatrix}
-2 & -4 & -2 \\
 0 &  0 &  0 \\
 2 &  4 &  2 \\
\end{bmatrix}_{[:,0,:]}, 
\begin{bmatrix}
-1 & -2 & -1 \\
 0 &  0 &  0 \\
 1 &  2 &  1 \\
\end{bmatrix}_{[:,1,:]}
\end{bmatrix}
    $}
\end{eqnarray}

\begin{eqnarray}
S_z =
\resizebox{7.5cm}{!}{$
\begin{bmatrix}
\begin{bmatrix}
-1 & -2 & -1 \\
-2 & -4 & -2 \\
-1 & -2 & -1 \\
\end{bmatrix}_{[:,:,-1]}, 
\begin{bmatrix}
 0 &  0 &  0 \\
 0 &  0 &  0 \\
 0 &  0 &  0 \\
\end{bmatrix}_{[:,:,0]}, 
\begin{bmatrix}
 1 &  2 &  1 \\
 2 &  4 &  2 \\
 1 &  2 &  1 \\
\end{bmatrix}_{[:,:,1]}
\end{bmatrix}\nonumber
    $}
\end{eqnarray}

Here, \( S_x \) represents the three 3$\times$3 kernels on the $yz$-plane at three different $x$ slices, measuring changes along the $x$ direction. Similarly, \( S_y \) consists of three 3$\times$3 kernels on the $xz$-plane at different $y$ slices to measure changes along the $y$ direction, and \( S_z \) represents the three 3$\times$3 kernels on the $xy$-plane at different $z$ slices to measure changes along the $z$ direction (see \fref{fig:sobel}).
By maintaining spatial relationships and gradient directionality using these three slices for each direction, Sobel kernels offer a more precise representation of 3D gradients. This enhanced spatial context and gradient sensitivity make them highly effective for detecting and analyzing volumetric changes in 3D medical images.

The gradient along each direction is computed by convolving each Sobel kernel, $S_x$, $S_y$, and $S_z$, with the input image $f(x,y,z)$: 
\begin{equation}
    \begin{gathered}
        \nabla_x(x,y,z) \approx S_x * f(x,y,z), \\
        \nabla_y(x,y,z) \approx S_y * f(x,y,z), \\
        \nabla_z(x,y,z) \approx S_z * f(x,y,z)
    \end{gathered}
\end{equation}
where the discrete spatial convolution in 3D is defined as  
\begin{eqnarray}\label{Conv_op}
S*f(x, y, z) = \sum_{(a, b, c) \in S_m^3} S(a, b, c) \, f(x - a, y - b, z - c)
\end{eqnarray}
Here,  $S_m^3\in \mathbb Z_+^3$ represents the set of all possible indices within the Sobel kernel, which in our case contains $3\times3\times3$ possible combinations.
The result of this operation is the gradient of the image along the direction defined by the Sobel kernel $S$.
We apply the convolution three times, once in each of the $x$, $y$, and $z$-directions, thus resulting in three gradients, namely $\nabla_x$, $\nabla_y$, and $\nabla_z$.

To obtain information about the atrophy present in a brain, we start by convolving the Sobel kernels $S$ with the registered image (i.e., moving image after registration), $M(x,y,z)$, as $S*M(x,y,z)$, which leads to $\nabla M_x$, $\nabla M_y$, and $\nabla M_z$.
We also perform the same operation on the fixed image, resulting in $\nabla F_x$, $\nabla F_y$, and $\nabla F_z$.

Next, we calculate the 3-dimensional angles $\theta_{M}$ and $\theta_F$ for the moving image and the fixed image, respectively, and compute their difference:
\begin{align}
\begin{split}
\Resize{8cm}{\theta_{\text{Diff}} = \Bigg| \tan^{-1} \bigg( \frac{\sqrt{\nabla M_x^2 + \nabla M_y^2}}{\nabla M_z} \bigg) -  \tan^{-1} \bigg( \frac{\sqrt{\nabla F_x^2 + \nabla F_y^2}}{\nabla F_z} \bigg) \Bigg|}
\end{split}
\end{align}

This angle difference indicates the structures' orientation disparity between the corresponding registered image (representing the subject's brain) and the fixed image (representing the MNI template or a healthy brain).
Therefore, it signifies abnormalities or atrophy present in the subject's brain, thus aiding in the detection and characterization of dementia or AD-related brain structural changes. As shown in \fref{pics}, SKAD provides an alternative approach to Jacobian maps for depicting and highlighting volumetric changes in the brain, and enables a more comprehensive understanding of structural alterations. 

\begin{figure}[!t]
    \centering
     {\subfloat[Reorientation]{\includegraphics[width=0.33\textwidth]{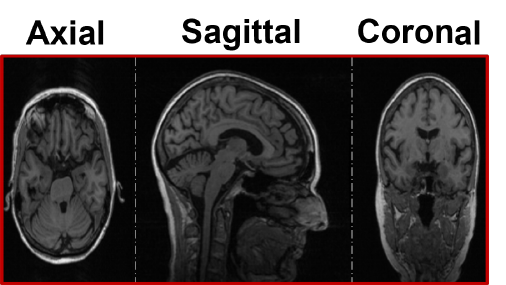}}} \hspace{1cm}
    {\subfloat[Bias Field Correction]{\includegraphics[width=0.33\textwidth]{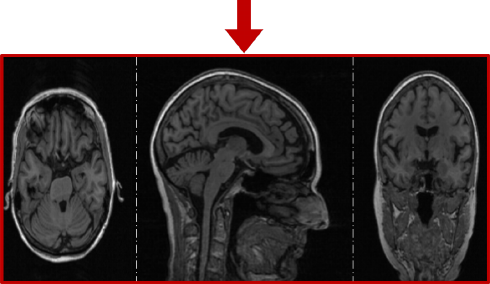}}}\hspace{1cm}
    {\subfloat[Brain Extraction]{\includegraphics[width=0.33\textwidth]{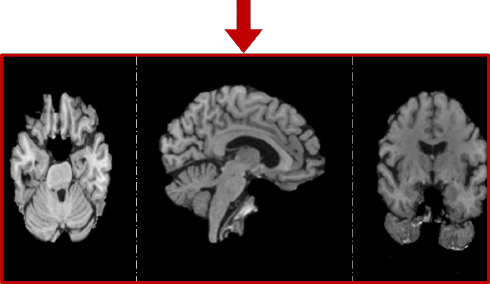}}}
\caption{The MRI preprocessing pipeline depicted across three orthogonal views of 3D MRI images.}  \label{preprocessing}
\end{figure}

\section{Evaluation}

\subsection{Datasets} 
{\bf Dataset Selection.} To evaluate our proposed approach, SKAD, we consider structural MRI images from two widely recognized and reputable organizations that collect data related to AD:
\begin{itemize} 
    \item { \bf Open Access Series of Imaging Studies { (OASIS)} \cite{lamontagne2019oasis}.} While they offer many MRI datasets, we specifically consider the OASIS-3 dataset in our evaluation. This dataset represents a recent advancement in the field, designed to deepen the understanding of dementia. It includes a diverse range of participants aged from 42 to 95 years, making it a valuable resource for studying various stages of cognitive decline.

    \item {\bf Alzheimer's Disease Neuroimaging Initiative {(ADNI)} \cite{weiner2010alzheimer}. } We also select the ADNI-I dataset from the available datasets for our evaluation, as it serves as a cornerstone in AD research. The ADNI-I dataset has been instrumental in facilitating robust comparisons with state-of-the-art methodologies. It encompasses a wide demographic, with participants aged from 40 to 89 years, and is extensively utilized in studies focused on the progression of AD.
\end{itemize}
Both datasets encompass a broad variety of subjects and imaging conditions, which enhances the robustness of our analyses. MRI is highly regarded in AD research due to its reliability as an indicator of disease progression. Structural MRI measures, which reflect brain atrophy, are recognized as robust markers of neurodegeneration \cite{frisoni2010clinical}. To maintain consistency across both datasets, we downloaded the raw data and applied identical preprocessing steps. This ensures that our evaluations are based on comparable data, enhancing the validity of our findings.

{\bf Data Labeling, Balancing, and Augmentation.} 
To label the datasets, we employ the clinical dementia rating (CDR) score to categorize the data into {\em four} distinct groups. The CDR is a widely used tool in clinical research and practice for assessing the severity of dementia and cognitive impairment. The CDR scale ranges from 0 to 3, defined as follows:
\begin{itemize} 
\item When the CDR score is 0, it indicates that the individual is cognitively normal (CN). 
\item When the score is 0.5, it indicates very mild impairment or questionable dementia. It is also known as mild {em cognitive impairment (MCI).}
\item When the score is 1, it indicates mild dementia (MLD).
\item When the score is 2, it indicates moderate dementia (MOD).
\item When the score is 3, it indicates severe dementia (SEV).
\end{itemize}
Given the limited number of subjects in the SEV category, we combine the MOD and SEV classes into a single category, designated as SEV. This results in four final classes: MCI, MLD, SEV, and CN.

To minimize the risk of overfitting, which can arise when subjects are included in both the training and testing sets \cite{altay2021preclinical}, we ensure that the same individuals do not appear in both datasets. Given that many subjects undergo multiple MRI sessions, this approach allows us to create a subject-disjoint dataset.

To address the class imbalance, we apply data augmentation to the training set, employing a method similar to that described in \cite{de2023explainable}. That is because pathological datasets often suffer from imbalances due to challenges in collecting equal amounts of data across all classification groups. To mitigate this, we randomly apply one of the following transformations to each training image:
\begin{itemize}
    \item \textbf{Rotate}: Randomly rotates the 3D volume by an angle between \([-10^\circ, +10^\circ]\).
    \item \textbf{Translate}: Shifts the volume by a range of \([-5, +5]\) units along each axis.
    \item \textbf{Roto-translate}: Combines both rotation and translation.
    \item \textbf{Zoom-in}: Applies a zoom factor ranging from 1.1 to 1.3.
\end{itemize}
We also implement anti-aliasing to smooth the transformations and resize the images to their original input dimensions. This augmentation process continues until the minority classes have a number of images comparable to that of the largest class, thereby increasing variability and reducing the risk of overfitting.

After augmentation, the class distributions for the ADNI dataset were rebalanced from (133, 54, 10, 54) to (133, 133, 133, 133) for the MCI, MLD, SEV, and CN classes, respectively. For the OASIS dataset, the classes were rebalanced from (183, 93, 20, 1233) to (1233, 1233, 1233, 1233).

{\bf Data Preprocessing.} Preprocessing is imperative in medical imaging registration, as discussed in \sref{sec:reg}. Our MRI image preprocessing pipeline, illustrated in \fref{preprocessing}, consists of three essential steps:
\begin{itemize}
    \item {\bf Reorientation.} During image acquisition, different subjects often exhibit varying orientations. To achieve uniformity for inspection and training, all images are reoriented to align with a standard template. We utilize the \texttt{fslreorient2std} tool within FSL \cite{jenkinson2012fsl}, which applies 90, 180, or 270-degree rotations along different axes to align the images while preserving their positional integrity.

    \item {\bf  Bias field correction.} MRI images frequently encounter a common issue known as {\em bias field} or intensity inhomogeneity, which results in non-uniform signal intensity across the image and can affect the accuracy of analysis and interpretation. To address this, we perform bias field correction using the ANTs algorithm \cite{avants2009advanced}, a robust image processing tool designed for various neuroimaging tasks.

    \item {\bf Brain extraction.} Brain extraction, or {\em skull stripping}, involves removing non-brain tissues from the MRI images to isolate the brain for further analysis, enhancing visualization and facilitating standardization. We employ the Brain Extraction Tool (BET) \cite{smith2002fast} for this step. While BET may occasionally fail to accurately delineate all non-brain material or inadvertently remove small brain regions, it is generally acceptable for the subsequent registration task.
\end{itemize}

\begin{table}[!t]
\centering
\setlength{\tabcolsep}{0.75em}  
\renewcommand{\arraystretch}{1.3}  
\centering
\caption{Computational Complexity Comparison}
\resizebox{.75\linewidth}{!}{%
\begin{tabular}{c|c|c}
 \toprule
\textbf{Function} & \textbf{Total Time (ms)} & \textbf{Memory Usage (MiB)}\\ 
\midrule
Registration  & 12972 & 1364.3 \\
\hline
Jacobian Map  & 584 & 1267.3 \\
\midrule \hline
 \cellcolor{red!10} SKAD  &  \cellcolor{red!10} \textbf{93} & \cellcolor{red!10} \textbf{1199.7} \\
 \bottomrule
\end{tabular}
}\label{comp}
\end{table}

\subsection{Classification Model}
For classification, we selected the VGG-16 model, known for its high accuracy across various image classification tasks, including those in medical imaging \cite{jain2021deep, ismael2021deep}. The pre-trained VGG-16, which has been trained on over a million images, captures general image features that can be fine-tuned for specific MRI analysis tasks, even with relatively small datasets. This is particularly beneficial in medical imaging, where large labeled datasets are often limited. VGG-16’s consistent use of 3x3 convolutional filters throughout the network simplifies its architecture, making it effective and easily adaptable for 3D MRI analysis. 

\subsection{Computational Analysis}\label{sec:comp}

{\bf Time and Memory}. \tref{comp} provides a breakdown of computational complexity for a 3D MRI image of size (182, 218, 182), using registration, jacobian map, and SKAD, along with details on the total time taken in milliseconds (ms) and memory usage in Mebibytes (MiB). Two profiling tools, memory profiler \cite{memory_profiler} and cProfile Python packages, were used to calculate these numbers. It's noteworthy to mention that the registration step is already computationally burdensome, and having less burdensome options after registration could be more efficient. We can see that SKAD  has the lowest total time (93 ms) and the lowest memory usage (1199.7 MiB).

\begin{figure}[t]
  \centering
 \includegraphics[width=\linewidth]{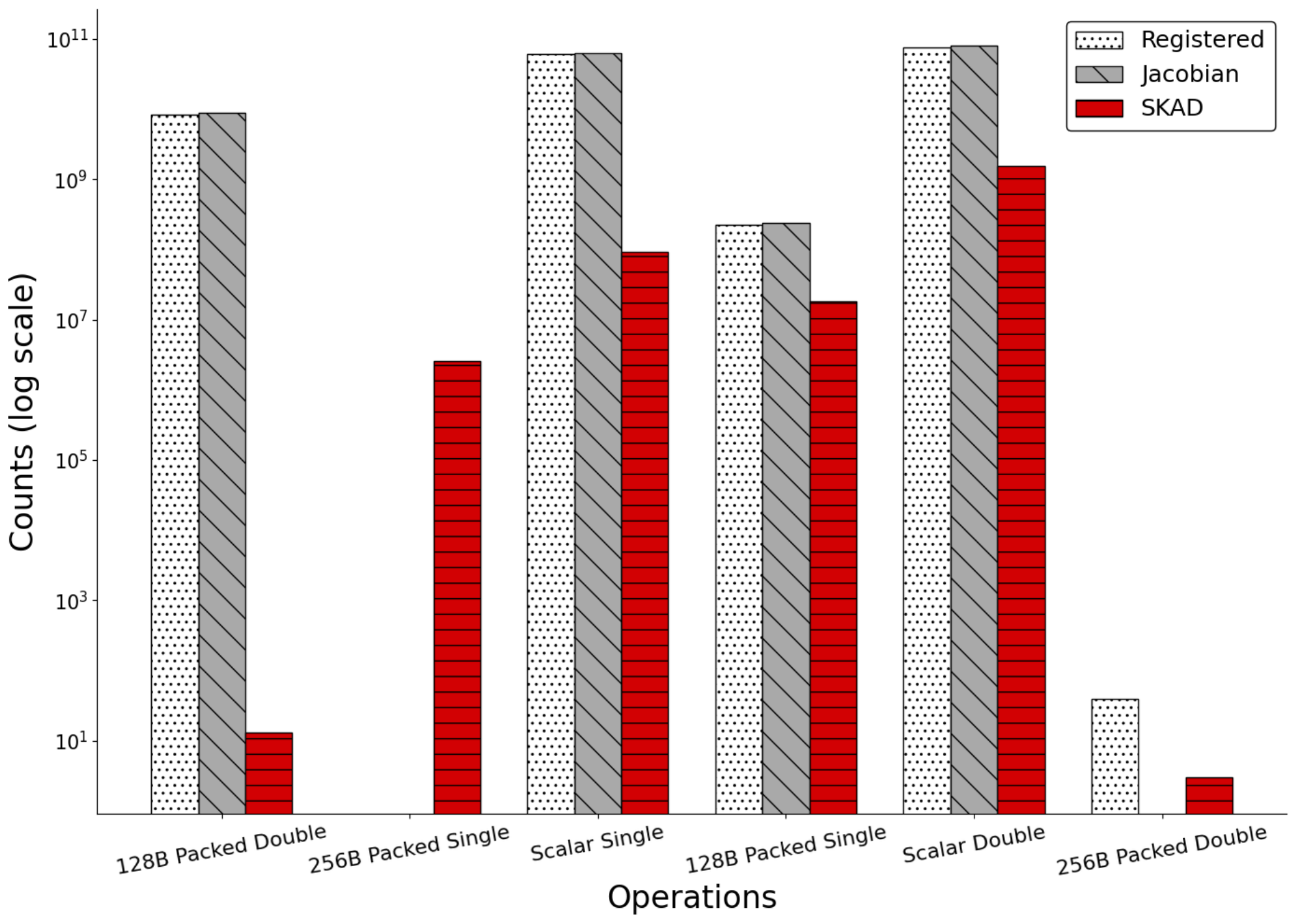}
  \caption{Comparison of Floating-Point Operations (note the log scale).}
  \label{flops}
\end{figure}

\begin{table*}[t]
\centering
\setlength{\tabcolsep}{1.2em}  
\renewcommand{\arraystretch}{1.5}  
\caption{Comparison of SKAD with state-of-the-art techniques which use Jacobian Maps}
\begin{tabular}{c|c|c|c|c|c|c}
\toprule
\textbf{Methods}& \textbf{Dataset} & \textbf{Model} & \textbf{Model Input} & \textbf{ACC} & \textbf{SENS} & \textbf{SPEC}  \\
 \midrule
\hline
Mustafa et al. 2024~\cite{mustafa2024unmasking}& OASIS-3  & Jacobian-Guided CNN & Registered CT and MRI  & 95.37 & 93.5 &  93.5  \\ 
\hline
Abbas et al.2023~\cite{abbas2023transformed} & ADNI & CNN  & Jacobian MRI & 96.6 & 97.8 & 95.9 \\
\hline
Mustafa et al. 2023~\cite{mustafa2023diagnosing}& OASIS-3  & RF and CNN & Jacobian MRI and CT & 98.7  & 97.2 &  95.2  \\
\hline
Spasov et al. 2019~\cite{spasov2019parameter} & ADNI    & CNN &  MRI, Jacobian, and clinical data & 83 & 87.5 & 81   \\
 \midrule\hline
  \textbf{SKAD} &   ADNI &   VGG-16  &   SKAD MRI &   91.06 &   83.5 &   \textbf{94.3} \\ 
\hline
  \textbf{SKAD} &   OASIS &   VGG-16  &   SKAD MRI &   \textbf{94.3} &  84.6 &   93.9  \\ 
\bottomrule
\end{tabular}
\label{lit}
\end{table*}

\begin{table*}[!t]
\setlength{\tabcolsep}{1em}
\centering
\renewcommand{\arraystretch}{1.5}
\caption{Ablation studies comparing model performance across registered images, Jacobian maps, and SKAD}
\resizebox{.95\linewidth}{!}{%
\begin{tabular}{l|l|c|cccc|c|cccc|c|ccccc}
\toprule
\multirow{2.2}{*}{Data} & \multirow{2.2}{*}{Input Type} & \multicolumn{5}{c|}{Sensitivity (\%)} & \multicolumn{5}{c|}{Specificity (\%)} & \multicolumn{5}{c}{Accuracy (\%)} \\
\cline{3-7} \cline{8-12} \cline{13-17}
 &  & \textbf{Avg} & CN & MCI & MLD & SEV & \textbf{Avg} & CN & MCI & MLD & SEV & \textbf{Avg} & CN & MCI & MLD & SEV \\
 \midrule
\hline
\multirow{3}{*}{ADNI} & \textbf{Registered} & 74.4 & 68.5 & 77.2 & 69.9 & 82.0 & 91.4 & 89.4 & 89.3 & 93.8 & 93.1 & 87.1 & 84.0 & 86.2 & 89.3 & 88.9 \\
\cline{2-17}
 & \textbf{Jacobian} & 87.8 & 80.3 & 75.1 & 96.7 & 99.0 & 96.0 & 94.7 & 92.2 & 97.7 & 99.4 & 94.0 & 90.7 & 88.0 & 97.8 & 99.6 \\
\cline{2-17}
 & \textbf{SKAD} & 83.5 &  78.10 &  88.6 &  78.42 &  89.04 &  94.3 &  91.65 &  89.86 &  95.80 &  100.0 &   \textbf{91.06} &  88.65 &  84.23 &  91.94 &  99.42 \\
 \midrule\hline
\multirow{3}{*}{OASIS} & \textbf{Registered} & 77.7 & 37.5 & 94.2 & 81.3 & 97.9 & 92.4 & 98.7 & 74.1 & 97.9 & 99.0 & 88.7 & 82.9 & 78.3 & 94.2 & 99.6 \\
\cline{2-17}
 & \textbf{Jacobian} & 83.9 & 64.4 & 84.4 & 87.8 & 99.0 & 92.8 & 85.7 & 98.6 & 87.8 & 99.0 & 89.4 & 85.0 & 85.0 & 87.9 & 99.8 \\
\cline{2-17}
 &  \textbf{SKAD} &   \textbf{84.6} &  67.26 &  84.13 &  94.54 &  92.67 &  \textbf{93.9} &  94.46 &  98.50 &  82.99 &  99.64 &   \textbf{94.3} &  92.35 &  94.43 &  91.64 &  98.88 \\
\bottomrule
\end{tabular}}
\label{ablation}
\end{table*}

\begin{figure*}[!t]
    \centering
     {\subfloat[ADNI performance metric.]{\includegraphics[width=0.49\textwidth]{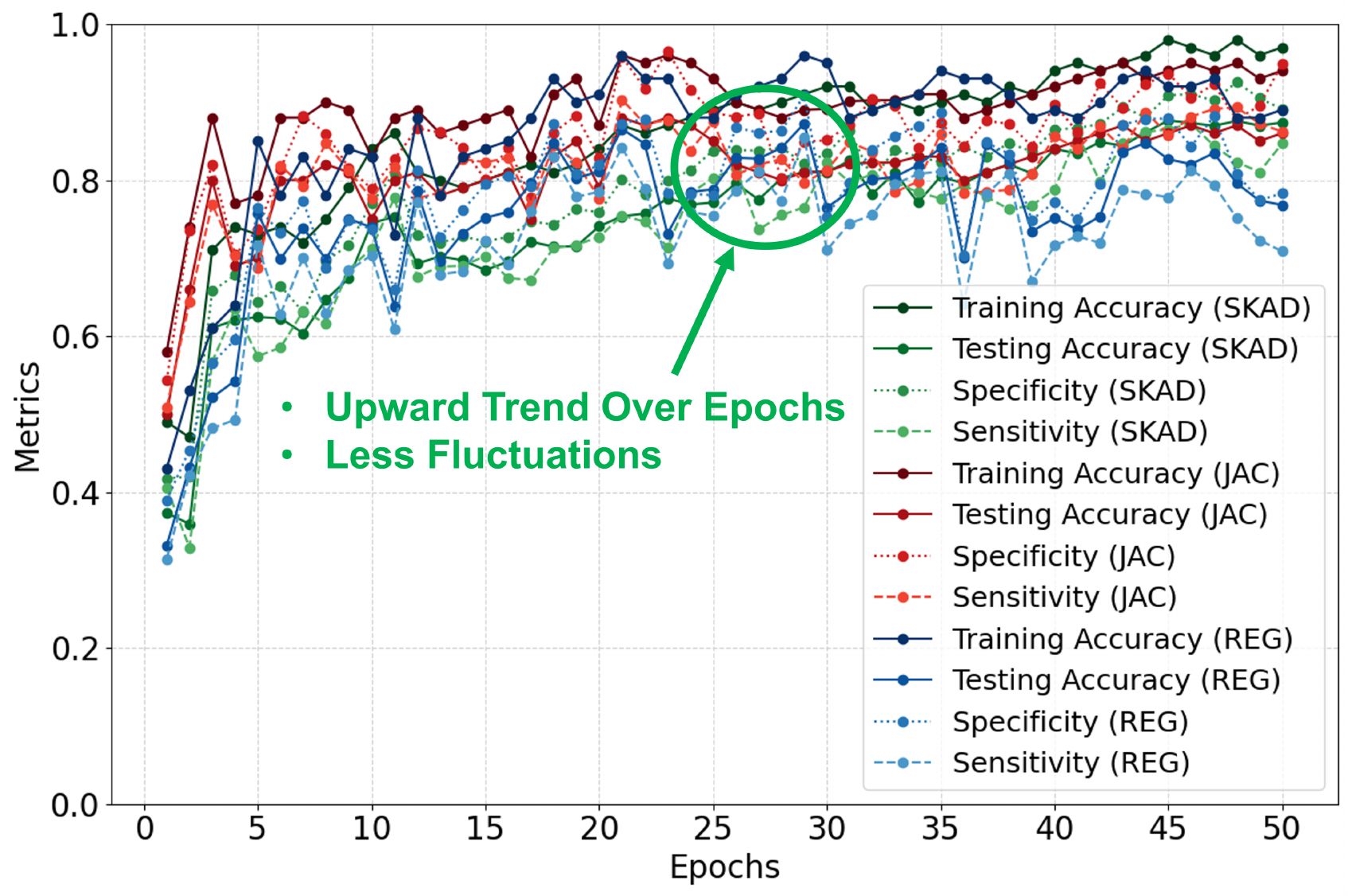}}}
     \hfill
    {\subfloat[OASIS performance metrics.]{\includegraphics[width=0.49\textwidth]{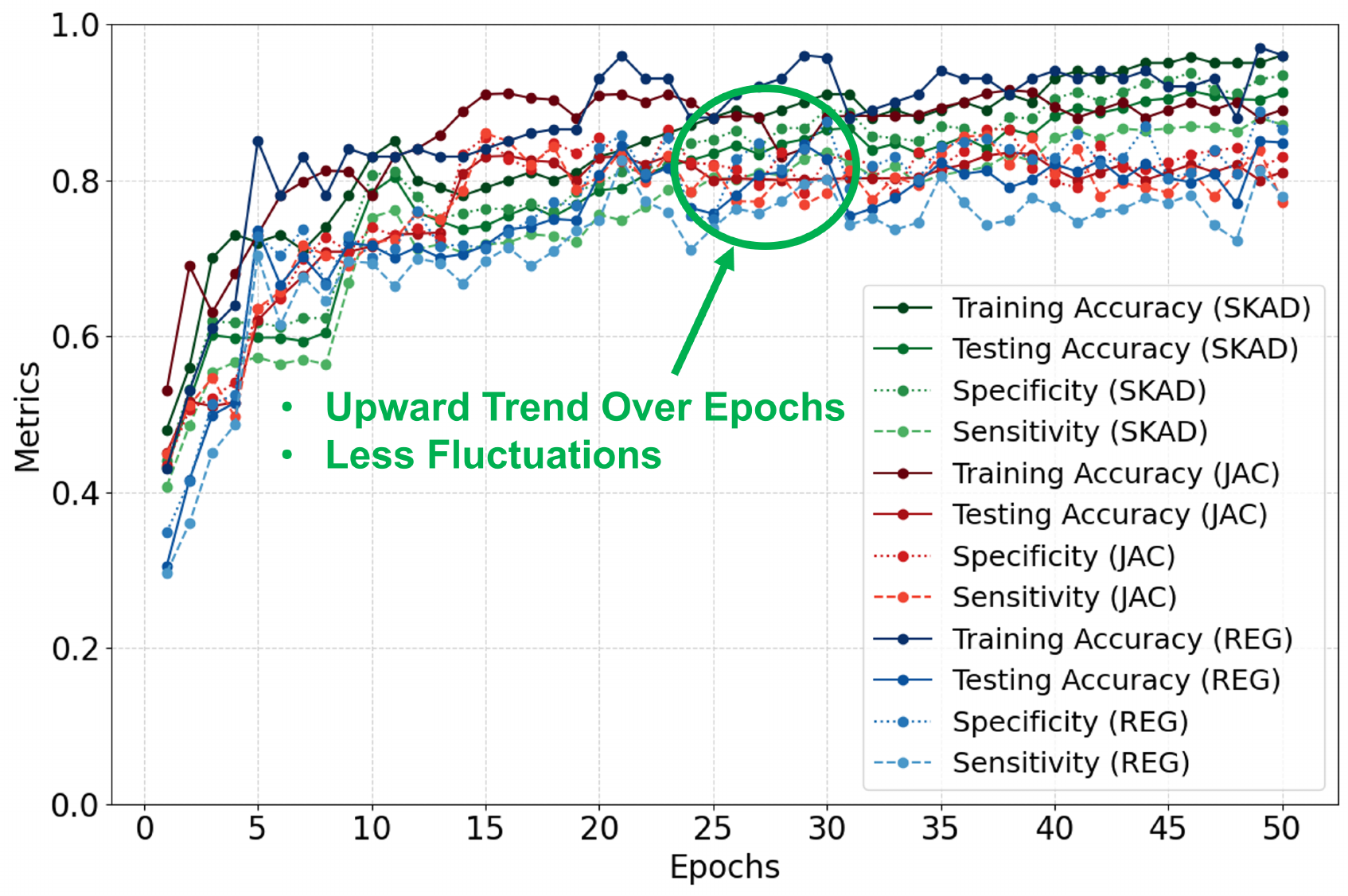}}}
    
\caption{Multiple performance metrics for 3 methods on ADIN and OASIS datasets over 50 epochs. The text annotation describes SKAD.}\label{performance} 
\end{figure*}

{\bf Floating-Point Operations}. Common floating-point number operations include addition, subtraction, multiplication, and division. We measure the number of floating-point operations using {\tt perf-stat} on Linux~\cite{kerrisk2010linux} and compare registration, Jacobian, and SKAD. The results are summarized in \fref{flops} for six operations: 1) \textit{128b Packed Double} represents the operations involving 128-bit packed double-precision floating-data, where ``Packed'' means that multiple values are processed together in a single instruction; 2) \textit{256b Packed Single} refers to the operations involving 256-bit packed single-precision floating-data; 3) \textit{Scalar Single} is operations on individual single-precision floating-point numbers; unlike packed operations, it deals with one value at a time; 4) \textit{128b Packed Single} operations use 128-bit packed single-precision floating-point data, where multiple single-precision values are handled simultaneously within one 128-bit width instruction; 5) \textit{Scalar Double} encompasses operations on individual double-precision floating-point numbers, which deals with one double-precision value at a time; 6) \textit{256b Packed Double} denotes operations with 256-bit packed double-precision floating-point data, processing multiple double-precision values in a single 256-bit width instruction.

As shown in \fref{flops}, SKAD requires a lower number of floating point operations than other methods in almost all 6 categories, indicating better computational efficiency and resource management. The only exceptions appear in the categories of 256b packed single {and double} where the reason might be that SKAD increases throughput by processing larger data sets at once, leading to higher counts in this category. However, when considering the sum across all the categories, the FP operations are approximately 153.02 billion (Registered), 146.01 billion (Jacobian), and 1.66 billion (SKAD). This remarkable improvement in computational efficiency achieved by SKAD can be attributed to several key factors:
\begin{itemize}
    \item {\bf Simplified Calculation:} SKAD uses Sobel kernels, which are simpler convolution operations than the complex derivative calculations required by Jacobian maps.
    \item {\bf Dimensionality Reduction:} Jacobian maps require calculating a full 3x3 matrix for each voxel, whereas SKAD only needs a single angle calculation, which significantly reduces the computational complexity.
    \item {\bf Optimized Convolution:} Sobel kernels can be efficiently implemented using optimized convolution operations, which are well-supported by modern hardware (e.g., NVDIA GPUs) and libraries.
    \item {\bf Single-Precision Optimization:} SKAD predominantly relies on single-precision operations, thus minimizing the use of computationally expensive double-precision operations (128B Packed Double, 256B Packed Double).
\end{itemize}

In summary, SKAD's lower computational requirements make it better suited for large-scale image analyses involving thousands of scans. Moreover, its faster processing would facilitate its integration into clinical workflows and thus enable more timely analysis of brain images.

{\bf State-of-the-Art Comparison}. 
\tref{lit} gives an overview of other works that also used Jacobian maps in their algorithms. SKAD has accuracies of 91.06\% (ADNI dataset) and 94.3\% (OASIS dataset). While these numbers may be slightly lower than some other works like  \cite{mustafa2023diagnosing}, the results are still generally strong. The specificity for SKAD is 94.3\% (ADNI dataset) and 93.9\% (OASIS dataset). These values are relatively high, indicating effectiveness in identifying true negatives. SKAD is competitive with existing state-of-the-art methods, demonstrating strong performance in terms of accuracy and specificity, but with slightly lower sensitivity. It is a unique input method with a focus on decreased computational complexity and offers insights into different aspects of MRI-based analysis. Given this context, SKAD can be positioned as a strong model with potential for further improvement in sensitivity, aiming to increase its overall effectiveness.

{\bf Ablation Study}.  Figure \ref{performance} demonstrates the performance of the training of the CNN model over 50 epochs for each dataset, ADNI and OASIS. We measure the performance using specificity (precision), sensitivity (recall), and accuracy. The figure demonstrates that using the Jacobian as input stabilizes the training more as seen from the red curves compared to the only registered input seen in blue. As for the SKAD (green curves), we can see that its performance is on par with Jacobian's performance and demonstrates the most stability in performance (especially with the ADNI dataset) compared to registered and Jacobian images. SKAD even demonstrates slightly higher performance than Jacobian on the OASIS dataset. 
In the ablation study \tref{ablation}, it is noteworthy that the Jacobian maps exhibit improved performance over registered images, indicating the effectiveness of Jacobian maps in enhancing the registration process. While SKAD may demonstrate higher performance in certain aspects, it largely performs on par with Jacobian maps while offering the advantage of reduced computational burden (see \sref{sec:comp}). Specifically, in the OASIS dataset, SKAD exhibits superior sensitivity for various classes, particularly in MCI and SEV categories, while maintaining comparable specificity and accuracy to Jacobian maps. However, in the ADNI dataset, the performance of both SKAD and Jacobian maps is generally lower, likely due to the smaller dataset size that we collected. Despite this, SKAD's ability to provide similar performance to Jacobian maps with reduced computational requirements underscores its potential as a viable alternative in neuroimaging analysis. 
\section{Conclusion and Future work}

This paper presents the Sobel Kernel Angle Difference (SKAD) method as a computationally-efficient alternative to Jacobian maps for characterizing brain volumetric changes in dementia diagnosis. Our findings demonstrate that SKAD delivers diagnostic results comparable to Jacobian maps while significantly reducing computational and resource demands. This advancement holds promise for enhancing the efficiency and accessibility of dementia diagnostics.

Tested on two diverse datasets, SKAD shows potential for broader applicability to other datasets, such as NACC and HCP. While primarily applied to Alzheimer’s disease and dementia, the method’s focus on brain atrophy and structural changes suggests it can be extended to other neurodegenerative diseases, such as Parkinson’s, provided relevant imaging data is available.

Future research will aim to enhance SKAD’s sensitivity by developing adaptive Sobel kernels capable of adjusting size and orientation based on local image characteristics, enabling finer gradient detection. Additionally, we plan to incorporate post-segmentation explanatory mechanisms and brain region labeling to make the results more interpretable and clinically actionable.

{\fontsize{9.5pt}{11pt}\selectfont
\bibliographystyle{ieee_fullname}
\bibliography{egbib}
}

\end{document}